\documentclass[prb,groupedaddress,showpacs,showemail]{revtex4}
\usepackage{graphicx}

\begin{document}

\title{Second-order transverse magnetic anisotropy induced 
by disorders in the single-molecule magnet Mn$_{12}$}
\author{Kyungwha Park$^{1,2,3,}$}\email{park@dave.nrl.navy.mil}
\author{Tunna Baruah$^{1,3}$}
\author{Noam Bernstein$^{1}$}
\author{Mark R. Pederson$^{1}$}
\affiliation{
$^1$Center for Computational Materials Science, Code 6390,
Naval Research Laboratory, Washington DC 20375 \\
$^2$Department of Electrical Engineering and Materials Science Research Center,
Howard University, Washington DC 20059 \\
$^3$ Department of Physics, Georgetown University, Washington DC 20057}
\date{\today}

\begin{abstract}
For the single-molecule magnet Mn$_{12}$, Cornia {\it et al.} recently 
proposed that solvent molecules may cause the quantum tunneling that 
requires a lower symmetry than S$_4$. However, magnetic quantum tunneling 
and electron paramagnetic resonance experiments suggested that the proposed 
theory may not correspond to the measurements.
In this regard, we consider positional disorder induced by the solvent molecules
and orientational disorder by the methyl groups of a Mn$_{12}$
molecule. We calculate, within density-functional theory, 
the second-order transverse magnetic anisotropy parameter $E$ 
and an easy-axis tilting angle $\theta$ induced by the positional 
disorder and the $E$ value by the orientational disorder.
We also calculate the local magnetic
anisotropy and the local easy axis for each inequivalent Mn site in 
different environments to investigate their effects on the global $E$ 
value. We find that the hydrogen bonding between a Mn$_{12}$ 
molecule and the solvent molecule is crucial to obtain 
a substantial $E$ value and that the $E$ value increases upon
geometry relaxations. Our calculations on relaxed geometries 
show that the largest calculated $E$ value
is 0.016~K and that the largest tilting angle is 0.5$^{\circ}$. 
Our largest $E$ value is comparable to experimental results but larger
than Cornia {\it et al.}'s.

\end{abstract}

\maketitle


\section{Introduction}

Single-molecule magnets (SMMs) are three-dimensional arrays of identical 
molecules. Each molecule has a large effective spin with a large magnetic 
anisotropy barrier, and responds to an external magnetic field as a 
nanoscale, single-domain magnetic particle. As such SMMs are ideal for 
high-density magnetic storage devices. The most extensively studied SMM is
[Mn$_{12}$O$_{12}$(CH$_3$COO)$_{16}$(H$_2$O)$_4$]
$\cdot$2(CH$_3$COOH)$\cdot$4(H$_2$O) (hereafter Mn$_{12}$).\cite{LIS80}
A single molecule of Mn$_{12}$ (Fig.~\ref{fig:Mn12geo}) has an effective 
ground-state spin of $S=10$ with a magnetic easy axis along the crystal 
$c$ axis (i.e., $z$ axis), and has S$_4$ symmetry.
Low-temperature magnetization hysteresis loop measurements on Mn$_{12}$ 
revealed steps at quantized magnetic fields, in which macroscopic
quantum tunneling (MQT) occurred between spin-up states and spin-down 
states through the magnetic anisotropy barrier.\cite{SESS93,FRIE96} 
Many competing models have been proposed to understand various features 
of the MQT: fourth-order transverse anisotropy,\cite{VILL94} thermally assisted 
quantum tunneling,\cite{FRIE96,GARA97} the Landau-Zener effect,\cite{RAED97} 
and dipolar interactions with dynamic hyperfine fields.\cite{PROK98}
However, the observation of quantized steps that are forbidden by the 
symmetry of a single Mn$_{12}$ molecule remains a long-standing puzzle.
It is imperative to fully understand the MQT because higher tunneling rates 
imply the loss of magnetically stored information, while tunneling is useful 
in quantum superposition of states crucial to quantum computing\cite{LEUE01-NAT}. 

A general effective single-spin Hamiltonian for SMMs is, to fourth order, 
\begin{eqnarray}
{\cal H}= D S_z^2 + E(S_x^2 - S_y^2) + C_1 S_z^4 
+ C_2 (S_+^4 + S_-^4) + {\cal H}^{\prime} + g\mu_B \vec{B} \cdot \vec{S} \:,
\label{eq:ham}
\end{eqnarray}
where $D$ is the uniaxial anisotropy parameter, $E$ is the second-order 
transverse anisotropy parameter, $S_z$ is a spin operator projected onto the 
easy axis, $C_1$ and $C_2$ are the fourfold longitudinal and transverse
anisotropy parameters, ${\cal H}^{\prime}$ contains all possible 
fourth-order terms that break S$_4$ symmetry, and the last term is 
the Zeeman interaction. The experimental parameter values are 
$D=-0.55$~K, $C_1=-0.00117$~K, and $C_2=\pm$2.88$\times$10$^{-5}$~K.\cite{BARR97}
The MQT is ascribed to the transverse terms in Eq.~(\ref{eq:ham}). 
According to the S$_4$ symmetry of a Mn$_{12}$ molecule, $E=0$ and 
${\cal H}^{\prime}=0$ so that the MQT is 
caused by $C_2$ alone and is allowed between states whose eigenvalues of $S_z$ 
differ by integer multiples of 4. However MQT is observed between states that
violate this rule. To understand these unexplained tunneling steps, 
it is necessary to include a lower symmetry than S$_4$. Transverse fields caused 
by internal fields (exchange, dipolar, and hyperfine fields) or field misalignment 
could induce these steps, but resulting tunneling rates will be too small compared 
to measured values\cite{MERT01}. Thus, a nonzero $E$ value is needed to 
expedite tunneling assuming that ${\cal H}^{\prime}$ is negligible. 
So far, two different theories have been proposed 
in this regard. Chudnovksy and Garanin\cite{CHUD01} proposed 
that possible dislocations in single crystals may provide a broad continuous 
distribution of $E$ and showed that a broad continuous range of $E$ values
is required to explain experiments. Detailed 
x-ray diffraction studies by Cornia {\it et al.}\cite{CORN02} 
suggested that disorder caused by acetic-acid (referred to as HAc) solvent 
molecules may break the S$_4$ symmetry and induce substantial locally varying
$E$ values. As discussed in Sec.~IIA, such disorder is expected on general
grounds. Recently, two completely different experiments\cite{BARC02,HILL03} have 
been performed to elucidate the sources of a low symmetry in Mn$_{12}$. 
Magnetic quantum tunneling measurements\cite{BARC02}
showed that tunneling splittings (between states whose eigenvalues
of $S_z$ do not differ by integer multiples of 4) have much narrower 
distributions than what the dislocation-induced theory suggested. 
They also showed that the largest $E$ value for low symmetry molecules 
is 0.01~K, which is twice larger than what Cornia 
{\it et al.}\cite{CORN02} calculated. Electron paramagnetic resonance measurements 
showed that an upper bound on the $E$ values is 0.014~K,\cite{HILL03} 
and that some molecules may have an appreciable easy-axis tilt away 
from the crystal $c$ axis.\cite{HILL03-2} 

The two different experiments revealed consistent results that may not
be fully explained by either of the two proposed theories. Thus, there is
a need for a new theory or a more refined theory that is consistent with
the experimental results. In this regard, we investigate the sources of 
a low symmetry in Mn$_{12}$ within density-functional theory (DFT). 
Specifically, we consider both positional disorder due to the HAc solvent 
molecules and orientational disorder caused by methyl-group rotations 
in Mn$_{12}$ molecules. We find that hydrogen bonding between the 
Mn$_{12}$ molecules and the solvent molecules is crucial in the second-order 
transverse anisotropy. We calculate the magnetic anisotropy parameters and 
the easy-axis tilting angles from the crystal $c$ axis for representative
configurations induced by the positional disorder that will be described later.
We find that the $E$ value increases upon geometry relaxation, and that
there exist relationships, due to symmetry, between the $E$ values for 
the representative configurations. We also calculate 
torsional barriers and the magnetic anisotropy parameters for 
configurations induced by the orientational disorder. Our calculated $E$ values
are comparable to the experimental data, while the easy-axis tilting
angles remain small. In Sec.~II, we 
describe our method and model for the positional disorder and for the
orientational disorder. In Sec.~III, we show our results and discuss 
their consequences. In Sec.~IV, we present our conclusion.

\section{Method and Model}

In our DFT calculations\cite{KOHN65}, we use spin-polarized 
all-electron Gaussian-orbital-based Naval Research
Laboratory Molecular Orbital Library (NRLMOL) \cite{PEDE90} 
within the Perdew-Burke-Ernzerhof (PBE) generalized-gradient 
approximation (GGA) for the exchange-correlation 
potential.\cite{PERD96} We consider the following simplified 
form of Mn$_{12}$: [Mn$_{12}$O$_{12}$(CH$_3$COO)$_8$(HCOO)$_{8}$(H$_2$O)$_4$]
$\cdot$ 2(CH$_3$COOH) where 8 formates (HCOO) substituted for 
8 acetates (CH$_3$COO, referred to as Ac) bridging Mn$^{4+}$ ions 
in the inner cubane and Mn$^{3+}$ ions in the outer crown 
(Fig.~\ref{fig:Mn12geo}). For simplicity, we do not 
include the water molecules of crystallization. Instead of a periodic 
structure, we examine an isolated Mn$_{12}$ molecule surrounded by  
four HAc solvent molecules. The zero-field magnetic anisotropy barrier 
for the $S=10$ ground-state manifold does not change much with our 
simplification nor with attachment of the solvent molecules. 
The total magnetic moment for the ground state was confirmed to be 
$20 \mu_{\mathrm B}$, which is in good agreement with experiment.  
It is common that the x-ray deduced experimental geometry has 
underestimated C-H and O-H bond lengths in comparison to standard 
hydrogen bond lengths, which results in large self-consistent 
forces on hydrogen atoms as large as an order 
of 1 hartree/bohr.\cite{PARK03-Mn4} Therefore,  
we use as an initial geometry the experimental geometry
with hydrogen positions corrected such that all C-H (O-H) bond lengths
become 1.1~\AA~(0.96~\AA).~ Then the initial geometry is self-consistently 
relaxed using NRLMOL until forces exerted on all atoms are small.  

\subsection{Positional disorder}
 
\begin{figure}
\includegraphics[angle=0,width=0.8\textwidth]{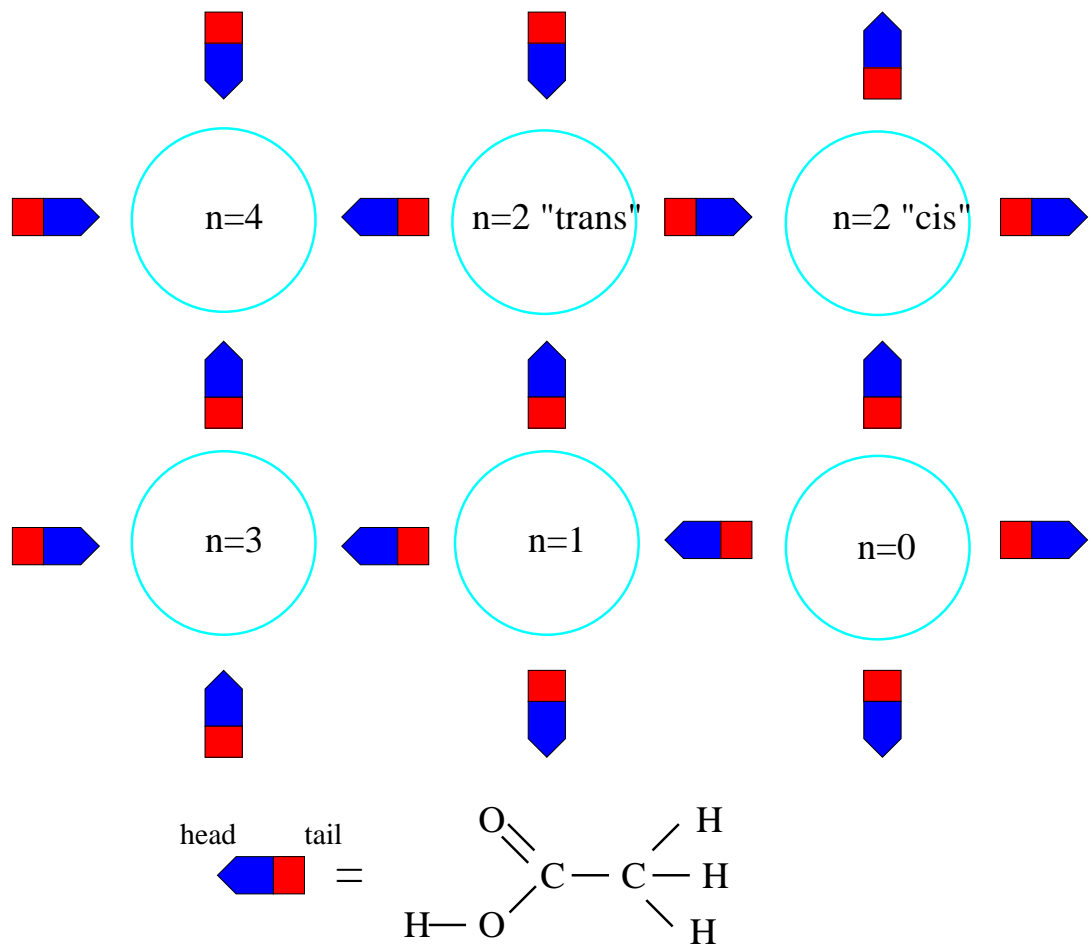}
\caption{Six distinct configurations of a Mn$_{12}$ molecule with
four HAc solvent molecules. The circles represent Mn$_{12}$ molecules.
$n$ is the number of HAc molecules whose heads point into a Mn$_{12}$
molecule. From the top leftmost, clockwise, the symmetry of each 
configuration is S$_4$, C$_2$, C$_1$, S$_4$, C$_1$, and C$_1$.}
\label{fig:isomers}
\end{figure}

A single Mn$_{12}$ molecule has four HAc solvent molecules, each 
of which is shared by two neighboring Mn$_{12}$ molecules. From
the perspective of the HAc molecule, there are two equivalent ways 
to place one HAc molecule between two Mn$_{12}$ molecules. That is, 
the methyl group (CH$_3$) in HAc can point
toward a Mn$_{12}$ molecule or away from it (Fig.~\ref{fig:isomers}). 
In the latter case, there is hydrogen bonding between the oxygen atom O(2) 
in the Mn$_{12}$ molecule and the oxygen atom O(1) in the solvent 
molecule (Fig.~\ref{fig:Mn12geo}). We refer to this configuration 
as {\it head} and the other as {\it tail} (Fig.~\ref{fig:isomers}). 
There are a total number of 16 ways to place four HAc molecules 
around a Mn$_{12}$ molecule. Only six of them are inequivalent by symmetry 
as illustrated in Fig.~\ref{fig:isomers} where $n$ denotes the number 
of the head configurations. The populations of the six configurations 
are shown in Table~\ref{table:1}. The configurations $n=1$, $n=2$ trans, 
$n=2$ cis, and $n=3$ break the S$_4$ symmetry, while $n=0$ and
$n=4$ preserve the S$_4$ symmetry. The three configurations, $n=1$, $n=2$ cis, 
and $n=3$, can cause easy-axis tilts from the crystal $c$ axis which
coincides with the $z$ axis, the easy axis without considering 
the positional disorder. 

\begin{figure}
\includegraphics[angle=0,width=0.8\textwidth]{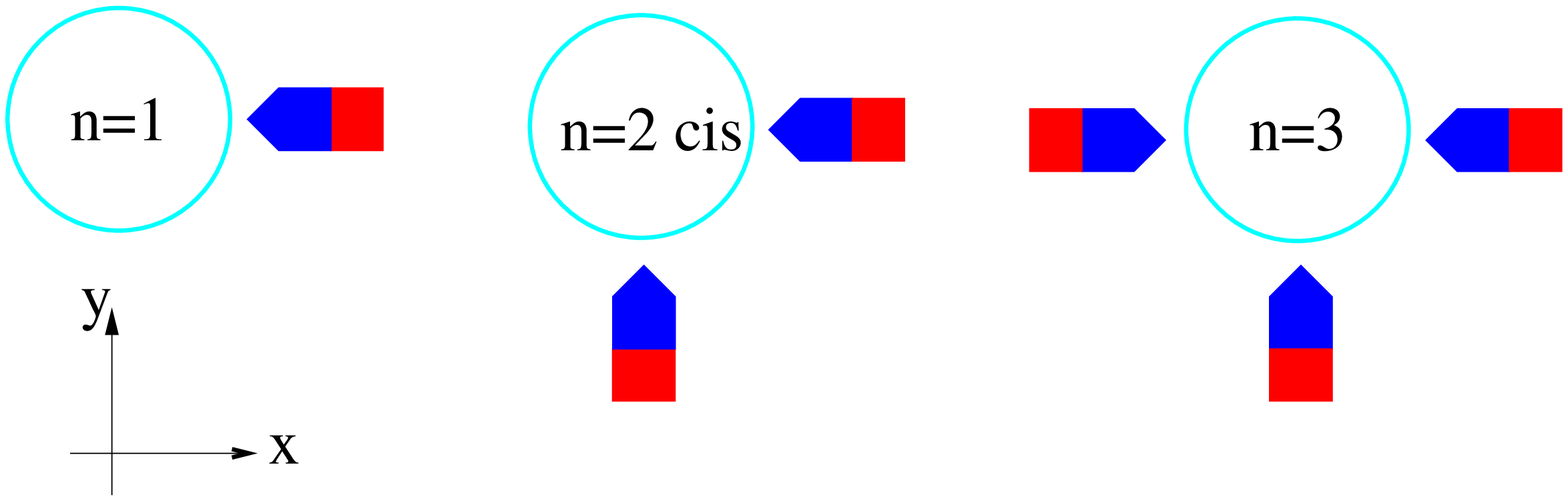}
\caption{Three head-only configurations used for 
Eqs.~(\ref{eq:sym3})-(\ref{eq:sym5}).}
\label{fig:symm}
\end{figure}

To examine if the head configuration affects the magnetic anisotropy
in the same way as the tail, we consider the following three 
configurations: (i) two tails only in a twofold symmetric way (trans), 
(ii) three heads and one tail, and (iii) three heads only. 
We calculate the $D$ and $E$ values for an initial geometry of each
configuration. We find that configuration (i) provides $D=-0.56$~K 
and $E=0.00053$~K and that configurations (ii) and (iii) provide 
the same parameter values of $D=-0.58$~K and $E=0.0035$~K.
These results imply that the tail configuration does not substantially
contribute to the second-order transverse anisotropy. 
It is thus sufficient to carry out DFT calculations on head-only 
configurations, which reduces significantly the computation time. 
Hereafter, unless specified, when we refer to the $n=1$ configuration,
it indicates only one head HAc added to a Mn$_{12}$ molecule
without three tail HAc molecules (see the leftmost figure in Fig.~\ref{fig:symm}). 
The same rule is applied to the rest of the configurations. 
We also find that both the transverse magnetic anisotropy
and the easy-axis tilting angle increase by a factor of 3 upon 
geometry relaxation.

Especially since we relax an isolated Mn$_{12}$ molecule with solvent instead of
a periodic structure, it is worthwhile to check if DFT provides
a correct hydrogen bond between two isolated atoms using
a simple example system. For this purpose, we carry out
DFT calculations for a water dimer. Using its fully relaxed geometry,
we obtain the distance between the two oxygen atoms, the O-H-O angle, 
and its binding energy. They are in excellent agreement with
experimental data\cite{ODUT80,CURT79,BENT80} as shown in 
Table~\ref{table:2}. This result confirms
that DFT captures the hydrogen bonding between Mn$_{12}$ molecules and
solvent molecules, and that relaxation of the initial geometries should 
provide physically refined geometries.

With a well relaxed geometry a second-order single-spin Hamiltonian 
is calculated considering spin-orbit coupling only.\cite{PEDE99} 
The Hamiltonian has a form of $\sum_{\mu,\nu=x,y,z} \gamma_{\mu \nu}S_{\mu} S_{\nu}$. 
Here $\gamma_{\mu \nu}$ is obtained from the calculated occupied and 
unoccupied states of the well relaxed geometry and contains the
information on the second-order magnetic anisotropy parameters 
and the principal axes of the magnetic anisotropy. By exploiting symmetry
arguments appropriate to the Mn$_{12}$ and the solvent molecules, 
we can explore all of the six different configurations (Fig.~\ref{fig:isomers}) 
using only the following 
two configurations. Once the single-spin Hamiltonians for the $n=0$ and 
$n=1$ configurations are calculated, those for the remaining four configurations can be 
obtained by rotations of coordinates. For instance, 
we can describe the Hamiltonian for $n=2$ trans (cis), ${\cal H}[n=2]$, 
in terms of those for $n=0$, ${\cal H}[n=0]$, and $n=1$, 
${\cal H}[n=1]$, using a rotation matrix $R$ (see Fig.~\ref{fig:symm}).
\begin{eqnarray}
{\cal H}[n=0] &\equiv& \sum_{\mu,\nu=x,y,z} \gamma^{(0)}_{\mu \nu}
S_{\mu} S_{\nu} \:, \: \: \: \: \: \: {\cal H}[n=1]
\equiv \sum_{\mu,\nu=x,y,z} \gamma^{(1)}_{\mu \nu} S_{\mu} S_{\nu} \:, 
\label{eq:sym1} \\
{\cal H}[n=2] &=& \sum_{\mu,\nu=x,y,z}
\gamma^{(1)}_{\mu \nu} [ S_{\mu} S_{\nu} + R(S_{\mu}) R(S_{\nu}) ]
- \sum_{\mu,\nu=x,y,z} \gamma^{(0)}_{\mu \nu} S_{\mu} S_{\nu} 
\nonumber \\
 & \equiv & \sum_{\mu, \nu=x,y,z} \gamma^{(2t),(2c)}_{\mu \nu} S_{\mu} S_{\nu} 
 \label{eq:sym2} \\
\gamma^{(2t)}_{xy}&=&2\gamma^{(1)}_{xy}, \: \: \: \gamma^{(2t)}_{xz}=0, \: \: \:
\gamma^{(2t)}_{yz}=0, 
\label{eq:sym3} \\
\gamma^{(2c)}_{xy}&=&0, \: \: \: \gamma^{(2c)}_{xz}=\gamma^{(1)}_{xz}-\gamma^{(1)}_{yz},
\: \: \: \gamma^{(2c)}_{yz}=\gamma^{(1)}_{xz}+\gamma^{(1)}_{yz} 
\label{eq:sym4} \\
\gamma^{(3)}_{xy}&=&\gamma^{(1)}_{xy}, \: \: \: \gamma^{(3)}_{xz}=-\gamma^{(1)}_{yz},
\: \: \: \gamma^{(3)}_{yz}=\gamma^{(1)}_{xz}
\label{eq:sym5}
\end{eqnarray}
where $\gamma^{(2t)}$, $\gamma^{(2c)}$, and $\gamma^{(3)}$ stand for the $\gamma$ 
matrices for the $n=2$ trans, $n=2$ cis, and $n=3$ configurations, respectively. 
Here we show only the off-diagonal elements of the $\gamma$ matrices. Since
$n=0$ has the S$_4$ symmetry, 
$\gamma^{(0)}_{xy}$=$\gamma^{(0)}_{xz}$=$\gamma^{(0)}_{yz}$=0. We have also 
verified this numerically.

\subsection{Orientational disorder}

The methyl group in Ac can rotate about the C-C single 
bond. During the rotation, the energy of the methyl group 
will change. The difference between the maximum and the minimum 
energy is called a torsional barrier. For an isolated Ac molecule, 
the torsional barrier is calculated to be about 320~K, which 
is comparable to the thermal energy at room temperature. 
So it is interesting to examine if the orientational disorder caused
by rotations of the methyl groups can induce the appreciable second-order 
transverse anisotropy. We consider rotations of the methyl groups in 
a Mn$_{12}$ molecule, and there are three inequivalent Ac sites.
The methyl groups in the Ac molecules bridging the inner Mn ions 
and the outer Mn ions will have less freedom to rotate and therefore 
a smaller effect on the transverse anisotropy. The methyl groups of 
type A (Fig.~\ref{fig:Mn12geo}) are close to the HAc solvent molecules 
so they also have less freedom to rotate. Thus we rotate the methyl 
groups of type B (in the Ac molecules that are free from the HAc molecules)
with the methyl groups of type A fixed. In this study, we do not allow
solvent-induced symmetry breaking and do not consider geometry
relaxation induced by methyl rotations. We consider an initial geometry 
of the $n=4$ configuration with several methyl groups of type B rotated, 
and calculate the torsional barriers and the magnetic anisotropy parameters. 

\section{Results}

\subsection{Positional disorder}

The well relaxed geometry for the $n=0$ configuration gives rise
to $D=-0.54$~K and $E=0$. The well relaxed geometry for the 
$n=1$ configuration provides $D=-0.54$~K, $E=0.008$~K, and 
the easy-axis tilting angle $\theta=0.4^{\circ}$. Using this result 
and Eqs.~(\ref{eq:sym1})-(\ref{eq:sym5}), we calculate $D$, $E$, and 
$\theta$ for the $n=2$ trans, $n=2$ cis, $n=3$, and $n=4$ 
configurations (Table~\ref{table:1}). The $D$ value does not vary 
much among the different configurations. (Note that $D=-0.55$~K without
solvent molecules.\cite{PEDE99}) The $E$ value for $n=1$ 
is the same as that for $n=3$ and is one half of that for $n=2$ 
trans. The $E$ value for $n=2$ cis is two orders of magnitude smaller 
that that for $n=1$. The largest $E$ value arises from $n=2$ trans,
while the largest $\theta$ comes from $n=2$ cis.
The $E$ values for the low-symmetry configurations 
are determined by all the off-diagonal elements of the $\gamma^{(1)}$ matrix, 
while the easy-axis tilts are due only to the nonzero $\gamma^{(1)}_{xz}$ and 
$\gamma^{(1)}_{yz}$ values [Eqs.~(\ref{eq:sym1})-(\ref{eq:sym5})]. 
For the positional disorder, the magnitudes of
all the off-diagonal elements of the $\gamma^{(1)}$ matrix are very small 
compared to those of the diagonal elements. Therefore, there is no 
appreciable easy-axis tilt and there is a relation between the $E$ 
values for different configurations. The $n=2$ cis configuration does 
not have a significant contribution to the transverse anisotropy.
If the magnitudes of the off-diagonal elements are significant, there is 
a substantial easy-axis tilt and the $E$ values for the distinctive 
configurations are not correlated. Additionally, the $E$ value for 
$n=2$ cis has the same order
of magnitude as those for the other low-symmetry configurations.
This occurs when extra electrons are donated to a Mn$_{12}$ molecule
since the locations of the extra electrons are intrinsically
dependent on donor locations.\cite{PARK03-prep} 
In comparison to Cornia {\it et al.}'s results,\cite{CORN02} our 
calculated $E$ values are larger by a factor of 3 and our easy-axis 
tilting angle is almost the same. In comparison to the experimental 
data,\cite{BARC02,HILL03,HILL03-2} 
our largest $E$ value is in good agreement but our largest easy-axis 
tilting angle is smaller than 1$^{\circ}$. 

\begin{figure}
\includegraphics[angle=0,width=0.8\textwidth]{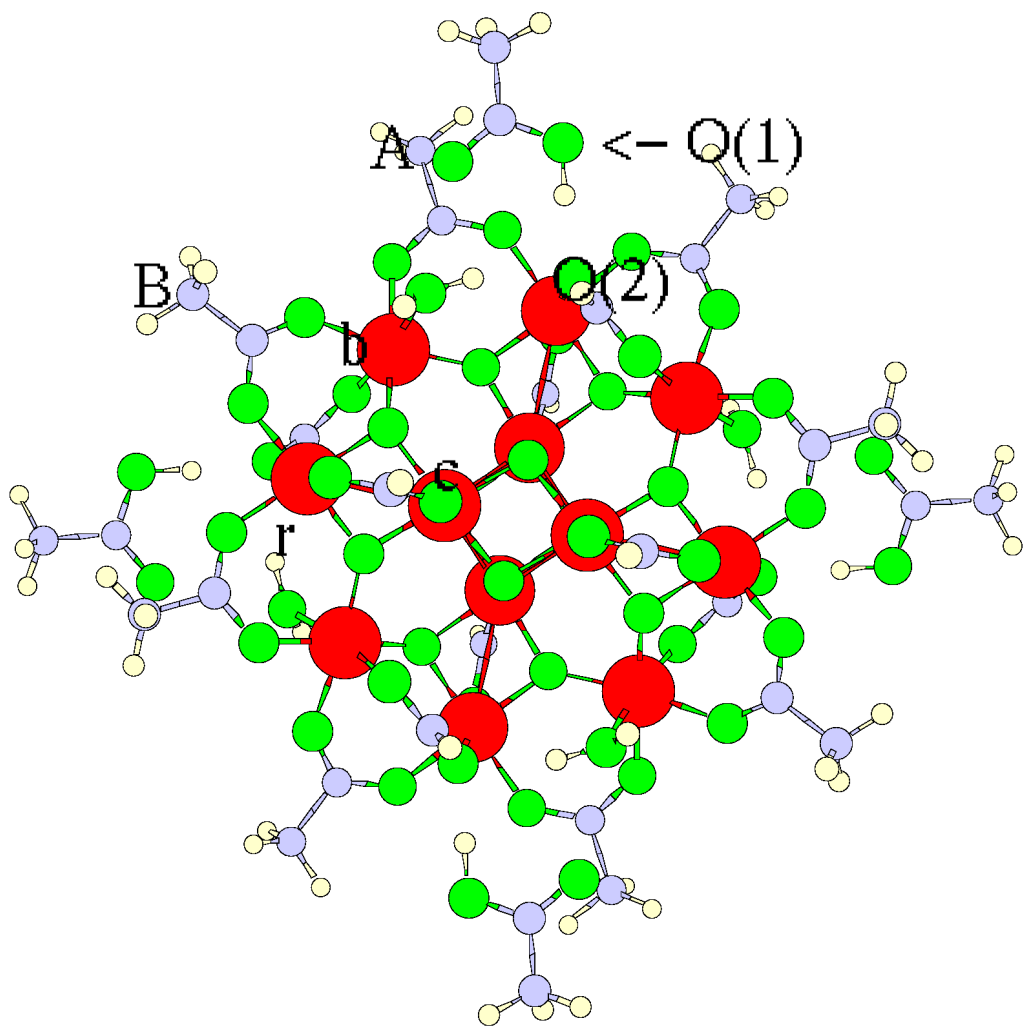}
\caption{A single Mn$_{12}$ molecule surrounded by four HAc
solvent molecules,
[Mn$_{12}$O$_{12}$(CH$_3$COO)$_8$(HCOO)$_{8}$(H$_2$O)$_4$]4(CH$_3$COOH).
The sizes of atoms decrease in the order of Mn, O, C, and H.
The four inner cubane Mn$^{4+}$ ions ($S=3/2$) are ferromagnetically 
coupled to each other and the eight outer crown Mn$^{3+}$ ($S=2$) ions
are ferromagnetically coupled. The Mn spins in the crown 
are antiparallel to those in the cubane, which results in the total effective 
ground-state spin of $S=10$. The easy axis of the molecule is normal to the page.
Two inequivalent acetates are labeled as ``A'' and ``B''. 
The hydrogen bond is formed between oxygen atoms O(1) and O(2).}
\label{fig:Mn12geo}
\end{figure}

To examine the effect of the positional disorder on the local magnetic 
anisotropy, we calculate the single-ion anisotropy parameters\cite{BARU02} 
and the local magnetic easy-axis tilts for the three inequivalent Mn sites
in two different environments. These three sites are the cubane Mn$^{4+}$
ion (at site c in Fig.~\ref{fig:Mn12geo}) and two Mn$^{3+}$ ions
(at sites r and b in Fig.~\ref{fig:Mn12geo}). The two environments 
correspond to (i) the head and (ii) tail configurations. 
For the environments (i) and (ii) we use a well relaxed $n=2$ trans (two heads 
and two tails) geometry. Before showing our results, we describe our method. 
The lowest-order energy shift $\Delta_2$ due to spin-orbit coupling 
$V_{LS}$ is\cite{PEDE99} 
\begin{eqnarray}
\Delta_2 &=& \sum_{\sigma, \sigma^{\prime}} \sum_{ij} \frac{ \langle
\psi_{i \sigma} | V_{LS} | \psi_{j \sigma^{\prime}} \rangle \langle
\psi_{j \sigma^{\prime}} | V_{LS} | \psi_{i \sigma} \rangle }
{\epsilon_{i \sigma} - \epsilon_{j \sigma^{\prime}}} \:,
\end{eqnarray}
where $\epsilon_{i \sigma}$ ($\epsilon_{j \sigma^{\prime}}$) denotes the energy 
of the $i$th occupied ($j$th unoccupied) unperturbed state whose wavefunction is
$|\psi_{i \sigma} \rangle$ ($|\psi_{j \sigma^{\prime}} \rangle$). The first
sum runs over spin-up and spin-down states and the second sum over 
all occupied and unoccupied states. A wavefunction of the system
is expanded in a sum of contributions associated with each atom,
$|\psi_{i \sigma} \rangle$$=$$\sum_{\alpha} |\psi^{\alpha}_{i \sigma} \rangle$
and $|\psi_{j \sigma^{\prime}} \rangle$$=$
$\sum_{\alpha} |\psi^{\alpha}_{j \sigma^{\prime}} \rangle$, where the
sums run over all atoms.
Thus, $\Delta_2$ can be written in terms of projected terms onto each atom
and mixed terms between different atoms as follows:
\begin{eqnarray}
\Delta_2&=& \sum_{\alpha} \sum_{\sigma, \sigma^{\prime}} \sum_{ij} \frac{ \langle
\psi^{\alpha}_{i \sigma} | V_{LS} | \psi^{\alpha}_{j \sigma^{\prime}} \rangle \langle
\psi^{\alpha}_{j \sigma^{\prime}} | V_{LS} | \psi^{\alpha}_{i \sigma} \rangle }
{\epsilon_{i \sigma} - \epsilon_{j \sigma^{\prime}}} 
+ \sum_{\alpha \beta \gamma \delta}
\sum_{\sigma, \sigma^{\prime}} \sum_{ij} \frac{ \langle \psi^{\alpha}_{i \sigma} |
V_{LS} | \psi^{\beta}_{j \sigma^{\prime}} \rangle 
\langle \psi^{\gamma}_{j \sigma^{\prime}} | V_{LS} | 
\psi^{\delta}_{i \sigma} \rangle } {\epsilon_{i \sigma} 
- \epsilon_{j \sigma^{\prime}}}
\label{eq:soc2}
\end{eqnarray}
where the first sum of the mixed term runs over all cases where at least two 
of the indices ${\alpha \beta \gamma \delta}$
are different. Based on the assumption that the contributions associated with
each atom, $|\psi_{i \sigma}^{\alpha} \rangle$, have small interatomic overlap, 
which is the case for Mn$_{12}$, the mixed terms are expected to 
be negligible. This is true considering the relative magnitude
of the Coulomb potential near an atom compared to the interstitial region.
Thus $\Delta_2$ is approximately equal to the tensor sum 
of the projected terms onto all atoms, the first term in Eq.~(\ref{eq:soc2}). 
Regardless of the types of the environments,
a cubane Mn$^{4+}$ ion has the local easy axis almost in the $xy$ plane 
(the local $E$ value is of the same order as the local $D$ value) and the two
crown Mn$^{3+}$ ions have the local easy axis close to the $z$ axis. Therefore 
the Mn$^{3+}$ ions contribute most of the total magnetic anisotropy. 
Since the Mn$^{4+}$ ions are far away from the solvent molecules, 
the local $E$ value for the head configuration is the same as that 
for the tail. For the two inequivalent Mn$^{3+}$ ions, the local $E$ value 
is sensitive to the environments, while the local $D$ value 
and the local easy-axis tilting angles are not sensitive to them. 
Especially for the Mn$^{3+}$ ions of type b, the local $E$ value
for the tail configuration is an order of magnitude smaller than 
that for the head. This large difference in the local $E$ values 
of the head and tail configurations leads to the large global $E$ value. 

We also calculate tunnel splitting $\Delta_{-10,4}$ between $M_s=-10$ 
and $M_s=4$, where $M_s$ is an eigenvalue of $S_z$ (Table~I) for the six 
distinctive configurations perturbatively and by exact diagonalization. 
We use our calculated values for $D$ and $E$ 
and the experimental values\cite{BARR97} for $C_1$ and $C_2$. 
The tunneling between $M_s=-10$ and $M_s=4$ is allowed due to 
the nonzero second-order transverse anisotropy but the magnitude of 
the tunnel splitting is mostly governed by the fourth-order 
transverse anisotropy. So the tunnel splitting is very dependent on 
the $C_2$ value but less on the $E$ value. As shown in Table~I, 
if we reduce the $C_2$ value by one half from the experimental value
while keeping other parameter values constant, 
the tunnel splitting is reduced by an order of magnitude. Thus, 
the experimental value of $\Delta_{-10,4}$[\onlinecite{MERT01}] 
may not be a good indicator of the magnitude of the $E$ value.

\subsection{Orientational disorder}

When the four methyl groups of type B are rotated relative to those of type A
in the $n=4$ configuration, we find that there are at least four local energy 
minima at the rotation angles of $-10^{\circ}$, $-25^{\circ}$, 0$^{\circ}$, 
and 30$^{\circ}$ where $0^{\circ}$ denotes the experimental geometry. 
Since the four methyl groups are rotated simultaneously, we have to divide by 
4 to obtain the torsional barriers per methyl group for each rotation angle. 
As shown in Table~\ref{table:4}, the torsional barriers at the three nonzero 
rotation angles are comparable to the thermal energy at room temperature. 
So we consider the three twofold symmetric configurations, which correspond
to $n=4$ with two methyl groups of type B across from each other rotated by 
$-10^{\circ}$, $-25^{\circ}$, and 30$^{\circ}$. Our calculations show
that the $E$ values for all three angles are on the order of 10$^{-4}$~K,
which is much smaller than those due to the positional disorder (Table~\ref{table:4}).

\section{Conclusion}

Considering the solvent positional disorder and the orientational disorder 
of the methyl groups in the Mn$_{12}$ molecules, we investigated, within
density-functional theory, the sources of the second-order transverse 
magnetic anisotropy in Mn$_{12}$. We found that only hydrogen-bonded solvent 
molecules contribute to the symmetry-breaking transverse anisotropy, and 
that geometry relaxation is necessary from the standpoint of quantitative
prediction. For the positional disorder, the large $E$ value is due to the large 
difference between the local $E$ values of the two different environments 
(the head and tail configurations). There are no significant easy-axis tilts from the $z$ 
axis for the low-symmetry configurations. Accordingly, the $E$ values for 
the different configurations are correlated. The orientational disorder 
contributes to the $E$ value much less than the positional disorder. 
Our largest calculated $E$ value from the positional disorder quantitatively 
agrees with the experimental data. However unresolved issues remain. 
That is, the experimental $E$ values suggest a {\it continuous} distribution 
rather than a discrete distribution that is predicted by the positional disorder.

\section*{Acknowledgments}
This work was supported in part by the ONR (Grant No. N000140211045), 
the NRL, the DoD HPCMPO and CHSSI program, and the W.M. Keck Foundation
(K.P.).

\clearpage

\begin{table}
\begin{center}
\caption{Populations, calculated magnetic anisotropy parameters ($D$ and $E$),
easy-axis tilting angles $\theta$, and tunnel splittings between
$M_s=-10$ and $M_s=4$, $\Delta_{-10,4}$, for six distinctive configurations
of an isolated Mn$_{12}$ molecule with four HAc solvent molecules.
See Fig.~\ref{fig:isomers}. For $\Delta_{-10,4}$[A], we use $C_1=-$0.00117~K
and $C_2$=2.88$\times$10$^{-5}$~K.\cite{BARR97} For $\Delta_{-10,4}$[B], 
we use $C_1=-$0.00117~K and $C_2$=1.44$\times$10$^{-5}$~K.}
\label{table:1}
\begin{ruledtabular}
\begin{tabular}{|c|c|c|c|c|c|c|}
configuration & population & $D$ (K) & $E$ (K) & $\theta$ (degree) 
& $\Delta_{-10,4}$[A] & $\Delta_{-10,4}$[B] \\ \hline
$n=0$ & $\frac{1}{16}$ & -0.54 & 0 & 0 & 0 & 0 \\ \hline
$n=4$ & $\frac{1}{16}$ & -0.56 & 0 & 0 & 0 & 0 \\ \hline
$n=1$ & $\frac{4}{16}$ & -0.54 & 0.008 & 0.4 
& 7.0 $\times$ 10$^{-7}$ & 7.8 $\times$ 10$^{-8}$ \\ \hline
$n=2$ cis & $\frac{4}{16}$ & -0.55 & 0.00002 & 0.5 
& 1.8 $\times$ 10$^{-9}$ & 4.4 $\times$ 10$^{-11}$ \\ \hline
$n=2$ trans & $\frac{2}{16}$ & -0.55 & 0.016 & 0 
& 1.0 $\times$ 10$^{-6}$ & 9.3 $\times$ 10$^{-8}$ \\ \hline 
$n=3$ & $\frac{4}{16}$ & -0.55 & 0.008 & 0.4 
& 6.7 $\times$ 10$^{-7}$ & 8.1 $\times$ 10$^{-8}$ \\  
\end{tabular}
\end{ruledtabular}
\end{center}
\end{table}

\begin{table}
\begin{center}
\caption{DFT calculated hydrogen bond length between the two oxygen atoms 
$R$(O-O), the bond angle (O-H-O), and its 
binding energy for a water dimer in comparison
to experimental data.\cite{ODUT80,CURT79,BENT80} }
\label{table:2}
\begin{ruledtabular}
\begin{tabular}{|c|c|c|c|}
 & R(O-O) (\AA)~ & $\angle$(O-H-O) (degree) & binding energy (eV) \\ \hline
DFT &  2.906 & 174 & 0.223 \\ \hline
Exp. & 2.946 [\onlinecite{ODUT80}] & 174 [\onlinecite{CURT79}] 
& 0.232 $\pm$ 0.030 [\onlinecite{BENT80}] \\
\end{tabular}
\end{ruledtabular}
\end{center}
\end{table}

\begin{table}
\begin{center}
\caption{Calculated local magnetic anisotropy parameters ($D$ and $E$), 
local easy-axis tilting angles $\delta$, and local anisotropy barriers 
for the three inequivalent Mn sites in different environments.
Here ''c'', ''r'', and ''b'' stand for the three sites shown in Fig.~\ref{fig:Mn12geo}.
''head'' and ''tail'' denote the head and tail configurations of the HAc
solvent molecules.}
\label{table:3}
\begin{ruledtabular}
\begin{tabular}{|c|c|c|c|c|}
site & $D$ (K) & $E$ (K) & $\delta$ (degree) & barrier (K) \\ \hline
Mn$^{4+}$(c, head)  & -0.50 & 0.18 & 101.5 & 0.72 \\ \hline
Mn$^{4+}$(c, tail) & -0.52 & 0.18 & 88.4 & 0.77 \\ \hline
Mn$^{3+}$(r, head)  & -2.44 & 0.037 & 11.8 & 9.6 \\ \hline
Mn$^{3+}$(r, tail) & -2.47 & 0.016 & 11.4 & 9.8 \\ \hline
Mn$^{3+}$(b, head)  & -2.54 & 0.23 & 34.5 & 9.2 \\ \hline
Mn$^{3+}$(b, tail) & -2.65 & 0.036 & 36.2 & 10.4 \\ 
\end{tabular}
\end{ruledtabular}
\end{center}
\end{table}

\begin{table}
\begin{center}
\caption{Calculated torsional barriers per methyl group
and magnetic anisotropy parameters due to rotations of two methyl groups of type
B (Fig.~\ref{fig:Mn12geo}) across from each other in the $n=4$ configuration.
The barriers are calculated from the closest local minimum configurations.}
\label{table:4}
\begin{ruledtabular}
\begin{tabular}{|c|c|c|c|}
 rotation angle (degree) & torsional barrier (K) & $D$ (K) & $E$ (K) \\ \hline
-10 & 316 & -0.57 & 0.00009 \\ \hline
-25 & 308 & -0.57 & 0.0002  \\ \hline
30  &  60 & -0.57 & 0.0001  \\ 
\end{tabular}
\end{ruledtabular}
\end{center}
\end{table}

\clearpage

\begin{thebibliography}{10}

\bibitem{LIS80}
T.\ Lis, Acta Crystallogr.\ B\ {\bf 36}, 2042 (1980).

\bibitem{SESS93}
R.\ Sessoli, D.\ Gatteschi, A.\ Caneschi, and M.A.\ Novak, Nature (London)
{\bf 365}, 141 (1993).

\bibitem{FRIE96}
J.~R.\ Friedman, M.~P.\ Sarachik, J.\ Tejada, and R.\ Ziolo, Phys. Rev. Lett.
{\bf 76}, 3830 (1996).

\bibitem{VILL94}
J.\ Villain, F.\ Hartman-Boutron, R.\ Sessoli, and A.\ Rettori,
Europhys.\ Lett.\ {\bf 27}, 159 (1994);
F.\ Hartmann-Boutron, P.\ Politi, and J.\ Villain,
Int.\ J.\ Mod.\ Phys.\ B {\bf 10}, 2577 (1996);
A.\ Fort, A.\ Rettori, J.\ Villain, D.\ Gatteschi, and R.\ Sessoli,
Phys.\ Rev.\ Lett.\ {\bf 80}, 612 (1998).

\bibitem{GARA97}
D.~A.\ Garanin and E.~M.\ Chudnovsky, Phys.\ Rev.\ B {\bf 56},
11 102 (1997); F.\ Luis, J.\ Bartolom{\'{e}}, and F.\ Fern{\'{a}}ndez,
{\it ibid.} {\bf 57}, 505 (1998); M.~N.\ Leuenberger and D.\ Loss,
Phys.\ Rev.\ B {\bf 61}, 1286 (2000).

\bibitem{RAED97}
H.\ De Raedt, S.\ Miyashita, K.\ Saito, D.\ Garci\'{a}-Pablos,
and N.\ Garc{\'{\i}}a, Phys.\ Rev.\ B {\bf 56}, 11 761 (1997);
V.~V.\ Dobrovitsky and A.~K.\ Zvezdin, Europhys.\ Lett.\
{\bf 38}, 377 (1997); ,
Phys.\ Rev.\ B {\bf 61}, 12 200 (2000).

\bibitem{PROK98}
N.~V.\ Prokof'ev and P.~C.~E.\ Stamp, Phys.\ Rev.\ Lett.\ {\bf 80}, 5794 
(1998); W.\ Wernsdorfer, T.\ Ohm, C.\ Sangregorio, R.\ Sessoli, D.\ Mailly,
and C.\ Paulsen, {\it ibid.} {\bf 82}, 3903 (1999).

\bibitem{LEUE01-NAT}
M.~N.\ Leuenberger and D.\ Loss, Nature (London) {\bf 410}, 789 (2001).

\bibitem{BARR97}
A.~L.\ Barra, D.\ Gatteschi, and R.\ Sessoli, 
Phys.\ Rev.\ B {\bf 56}, 8192 (1997).

\bibitem{MERT01}
K.~M.\ Mertes, Y.\ Suzuki, M.~P.\ Sarachik, Y.\ Paltiel, H.\ Shtrikman, 
E.\ Zeldov, E.\ Rumberger, D.~N.\ Hendrickson, G.\ Christou, 
Phys.\ Rev.\ Lett. {\bf 87}, 227205 (2001).

\bibitem{CHUD01}
E.~M.\ Chudnovsky and D.~A.\ Garanin, Phys. Rev. Lett. {\bf 87}, 187203 (2001);
Phys. Rev. B {\bf 65}, 094423 (2002).

\bibitem{CORN02}
A.\ Cornia, R.\ Sessoli, L.\ Sorace, D.\ Gatteschi, A.~L.\ Barra, 
and C.\ Daiguebonne, Phys. Rev. Lett. {\bf 89}, 257201 (2002).

\bibitem{BARC02}
E.\ del Barco, A.~D.\ Kent, E.~M.\ Rumberger, D.~N.\ Hendrickson, and
G.\ Christou, Europhys. Lett. {\bf 60}, 768 (2002);
E.\ del Barco, A.~D.\ Kent, E.~M.\ Rumberger, D.~N.\ Hendrickson, and
G.\ Christou, Phys.\ Rev.\ Lett.\ {\bf 91}, 047203 (2003).

\bibitem{HILL03}
S.\ Hill, R.~S.\ Edwards, S.~I.\ Jones, N.~S.\ Dalal, and J.~M.\ North,
Phys.\ Rev.\ Lett.\ {\bf 90}, 217204 (2003).

\bibitem{HILL03-2}
R.~S.\ Edwards, S.\ Hill, S.\ Maccagnano, N.~S.\ Dalal, and J.~M.\ North
(unpublished).

\bibitem{KOHN65}
W.\ Kohn and L.~J.\ Sham, Phy. Rev. {\bf 140}, A1133 (1965).

\bibitem{PEDE90}
M.~R.\ Pederson and K.~A.\ Jackson, Phy. Rev. B {\bf 41}, 7453 (1990);
K.~A.\ Jackson and M.~R.\ Pederson, {\it ibid.} {\bf 42}, 3276 (1990);
D.~V.\ Porezag, Ph.D. thesis, Chemnitz Technical Institute, 1997.

\bibitem{PERD96}
J.~P.\ Perdew, K.\ Burke, and M.\ Ernzerhof, Phys. Rev. Lett. {\bf 77},
3865 (1996).

\bibitem{PARK03-Mn4}
K.\ Park, M.~R.\ Pederson, S.~L.\ Richardson, N.\ Aliaga-Alcalde,
and G.\ Christou, Phys.\ Rev.\ B {\bf 68}, 020405(R) (2003).

\bibitem{ODUT80}
J.~A.\ Odutola and T.~R.\ Dyke, J.\ Chem.\ Phys.\ {\bf 72}, 5062 (1980).

\bibitem{CURT79}
L.\ A.\ Curtiss, D.~L.\ Frurip, and M.\ Blander,
J.\ Chem.\ Phys.\ {\bf 71}, 2703 (1979).

\bibitem{BENT80}
R.~M.\ Bentwood, A.~J.\ Barnes, and W.~J.~O.\ Thomas,
J.\ Mol.\ Spectrosc. {\bf 84}, 391 (1980).

\bibitem{PEDE99}
M.~R.\ Pederson and S.~N.\ Khanna, Phys. Rev. B {\bf 60}, 9566 (1999).

\bibitem{PARK03-prep}
K.\ Park, T.\ Baruah, and M.~R.\ Pederson (unpublished).

\bibitem{BARU02}
T.\ Baruah and M.~R.\ Pederson, Chem.\ Phys.\ Lett.\ {\bf 360}, 144 (2002).



\end{thebibliography}
\end{document}